\documentclass[12pt]{article}
\usepackage{amssymb}
\usepackage{amsfonts,amssymb,amsmath,amsthm}
\usepackage{epsfig}
\usepackage{slashed}
\usepackage{graphicx}
\usepackage{mathrsfs}
\usepackage{bbm}
\def\and{{\rm and}}

\def\b{\beta}

\def\om{\omega}
\def\p{\partial}
\def\th{\theta}
\def\ti{\tilde}

\def\D{\Delta}

\def\O{\Omega}

\begin{document}
\renewcommand{\thefootnote}{\fnsymbol{footnote}}
\begin{titlepage}

\vspace{10mm}
\begin{center}
{\Large\bf Quantum tunneling and spectroscopy of noncommutative inspired Kerr black hole}
\vspace{16mm}

{\large Yan-Gang Miao${}^{1,2,}$\footnote{\em E-mail: miaoyg@nankai.edu.cn},
Zhao Xue${}^{1,}$\footnote{\em E-mail: illidanpimbury@mail.nankai.edu.cn} and
Shao-Jun Zhang${}^{1,}$\footnote{\em E-mail: sjzhang@mail.nankai.edu.cn}

\vspace{6mm}
${}^{1}${\normalsize \em School of Physics, Nankai University, Tianjin 300071, China}

\vspace{3mm}
${}^{2}${\normalsize \em Kavli Institute for Theoretical Physics China, CAS, Beijing 100190, China}}
\end{center}

\vspace{10mm}
\centerline{{\bf{Abstract}}}
\vspace{6mm}

We discuss the thermodynamics of the noncommutative inspired Kerr black hole by means of a reformulated Hamilton-Jacobi method
and a dimensional reduction technique.
In order to investigate the effect of the angular momentum of the tunneling particle,
we calculate the wave function to the first order of the WKB ansatz. Then, using a density
matrix technique we derive the radiation spectrum from which the radiation temperature can be read out. Our results show that the radiation of
this noncommutative inspired black hole corresponds to a modified temperature which involves the effect of noncommutativity.
However, the angular momentum of the tunneling particle has no influence on the radiation temperature. Moreover, we analyze the entropy
spectrum and verify that its quantization is modified neither by the noncommutativity of spacetime nor by the quantum correction
of wave functions.
\vskip 20pt
PACS Number(s): 04.70.-s; 04.70.Dy; 11.10.Nx
\vskip 20pt
Keywords:
Quantum tunneling, Hawking temperature, noncommutative inspired black hole
\end{titlepage}

\newpage
\renewcommand{\thefootnote}{\arabic{footnote}}
\setcounter{footnote}{0}
\setcounter{page}{2}
\pagenumbering{arabic}

\section{Introduction}
In 1974, Hawking proved~\cite{Hawking} that black holes are not so black as their name pronounces.
Rather, they radiate energy continuously like a black body with temperature given by $T_{\rm H} = \frac{\hbar \kappa}{2\pi}$,
where $\kappa$ is the surface gravity of black holes. He made his calculation based on quantum field theory in curved spacetimes,
which was so technically involved that some other different methods appeared~\cite{Hartle1976} subsequently for the study of black hole radiation
called Hawking radiation now. However, none of the methods corresponds directly to the heuristic picture proposed by
Hawking that the source of radiation can be visualized as a tunneling process.

Since then, many attempts have been made to derive the radiation temperature and spectrum in terms of the tunneling picture. For instance,
Srivinasan and Padmanabhan proposed~\cite{Srinivasan} a
suggestion in accordance with the Hamilton-Jacobi equation, and shortly later, Parikh and Wilczek presented~\cite{Wilczek99} a more concise
way by means of the radial null geodesic of the tunneling particle. Note that the WKB ansatz is adopted semiclassically in the both methods which
reproduce Hawking's result but usually contain some deformation terms. For the
special case that the tunneling particle is described by the Klein-Gordon equation~\cite{Banerjee},
the wave function can be calculated exactly beyond the semiclassical approximation,
and therefore the deformation terms can be viewed physically.
In recent years the both methods have been applied to various kinds of
black holes~\cite{tunneling}, including noncommutative black holes~\cite{NC tunneling}.
However, for general cases one cannot obtain the exact wave function.

Nonetheless, Banerjee and Majhi provided~\cite{Banerjee09} a reformulated tunneling method for the spherically symmetric black holes.
They started with the Klein-Gordon
equation and simultaneously considered the WKB ansatz. Then, they calculated both the left and right moving wave functions inside or outside the horizon.
With the help of Kruskal coordinates, they made a connection between the wave function inside the horizon and the wave function outside the horizon.
After using the density matrix
technique they therefore obtained the emission spectrum. Their method was soon developed~\cite{Umetsu09} to study black holes with angular momenta.
Because of the no-hair theorem, this method can be extended to all the types of black holes. For more applications,
see some other recent papers~\cite{Banerjeeetal}.

Let us briefly recall the noncommutative spacetime and its related black holes, i.e. the noncommutative black holes and the noncommutative inspired ones.
Spacetime noncommutativity was first expected to resolve~\cite{Snyder} the divergent problem in quantum field theory,
but this goal was achieved by the renormalization
except for in quantum gravity. Later the noncommutativity
was expected to be an indispensable ingredient for quantization of gravity~\cite{DFR}.
Since the Seiberg and Witten's seminal work~\cite{Witten}, noncommutative theories have been appealing. Nowadays, the noncommutative theories are usually
regarded as a candidate at the energy level high up to the Planck scale. 
Until now,
quite a number of papers have been devoted to
the construction of quantum field theories and gravity on noncommutative spacetimes, for a review, see refs.~\cite{NCQFT,NCG}.
In classical gravity, the solutions of the Schwarzschild, Reissner-Nordstr\"{o}m and Kerr-Newman black holes
on noncommutative spacetimes have been given~\cite{Black Hole,Nicolini,Nicolini2010}.
Among the different techniques used to calculate noncommutative effects,
much attention has been paid to the star product approach~\cite{NCQFT}
and the coordinate coherent states approach~\cite{Smailagic and Spallucci}.
For the former, the noncommutative effect on both the matter and gravity is considered,
and therefore it is in general hard to find exact solutions to the noncommutative deformed Einstein equation.
Generally, one can only get perturbative solutions~\cite{Black Hole} to some order of the noncommutative parameters.
Only in some very special cases can the nonperturbative solutions be found, such as that discussed
in ref.~\cite{Twistedbh}. Based on the twisted noncommutative gravity theory~\cite{TDG}, the nonperturbative solutions are derived under some
specific compatibility conditions of metrics and twists. For the latter,
the main idea is that the point matter distribution should be depicted
by a Gaussian function rather than a Dirac delta function. In addition, an assumption must be
made that the Einstein's equation is unchanged when one searches for a black hole solution.
That is, the modification caused by the noncommutativity is only the deformation of the energy-momentum tensor.
Therefore the noncommutative black holes~\cite{Nicolini,Nicolini2010} induced by the coordinate coherent states approach are usually called the
noncommutative inspired ones. In some sense, the coordinate
coherent states approach might be regarded as an effective way at least from
a phenomenological viewpoint.

In this paper, we discuss the radiation of the noncommutative inspired Kerr black hole~\cite{Nicolini2010} by means of the reformulated tunneling
method~\cite{Banerjee09}.
Our purpose is to investigate how the angular momentum of the tunneling particle and
the noncommutativity of spacetimes affect the thermodynamics of the noncommutative inspired black hole.
We obtain some new results, that is, the angular momentum does not modify the radiation temperature but the noncommutativity does,
and significantly both the angular momentum and the noncommutativity do not modify the entropy quantum of the black hole.
It is not so surprising that the angular momentum does not contribute for  the noncommutative inspired Kerr black hole,
which can be understood~\cite{Robinson} by dealing with Hawking radiation
as an effective (1+1)-dimensional theory near the horizon for the ordinary black holes.
However, it is quite interesting that the noncommutativity of spacetime has no contribution to the entropy quantum.

The present paper is organized as follows.
In the next section,
we give a brief introduction to the noncommutative inspired Kerr black hole. In section 3, following the reformulated tunneling method, we provide the
relations between the left (right) moving wave function inside the horizon and the left (right) moving wave function outside the horizon.
In section 4, by using the density matrix technique
we obtain the emission spectra for both the bosons and fermions as tunneling particles.
In section 5, we calculate the entropy spectroscopy. Finally, we make a conclusion
in section 6. Throughout the paper, we adopt the geometrized units where $G=c=k_B=1$.

\section{Noncommutative inspired Kerr black hole}
In the so-called ``coordinate coherent states approach"~\cite{Smailagic and Spallucci},
a point mass $M$ is smeared over a region with width $\sqrt{\theta}$
and is depicted by a Gaussian function rather than a Dirac delta function as follows,
\begin{eqnarray}
\rho_{\th}(r)=\frac{M}{\left(4\pi\th\right)^{\frac{3}{2}}}e^{-\frac{r^2}{4\th}},
\end{eqnarray}
where $\th$ denotes the noncommutative parameter. Under the assumption that Einstein equation is untack,
one solution of the noncommutative inspired Kerr black hole with the matter source mentioned above has been found in ref.~\cite{Nicolini2010}
in terms of the magical Newman-Janis algorithm~\cite{NewmanJanis}. The line element in Boyer-Linquist coordinates is\footnote{The detailed analysis
on how to obtain the solution with the axial symmetry from the source with the spherical symmetry has been given in ref.~\cite{Nicolini2010}.}
\begin{eqnarray}
ds^2 &=& -\frac{\Delta-a^2\sin^2\vartheta}{\Sigma} dt^2 + \frac{\Sigma}{\Delta} dr^2
+ \Sigma d\vartheta^2-2a^2\sin^2\vartheta \left(1-\frac{\Delta-a^2\sin^2\vartheta}{\Sigma}\right) dt d\phi\nonumber\\
&&+\sin^2\vartheta\left[\Sigma+a^2\sin^2\vartheta\left(2-\frac{\Delta-a^2\sin^2\vartheta}{\Sigma}\right)\right] d\phi^2, \label{metric}
\end{eqnarray}
where $a\equiv J/M$, $J$ is the Komar angular momentum,
\begin{eqnarray}
\Delta(r)=r^2-2rm(r)+a^2,\qquad\Sigma(r,\vartheta)=r^2+a^2\cos^2\vartheta,
\end{eqnarray}
and
\begin{eqnarray}
m(r)=\frac{2 M}{\sqrt{\pi}} \gamma\left(\frac{3}{2},\frac{r^2}{4\th}\right).
\end{eqnarray}
The gamma function is defined as $\gamma (b,c)= \int_0^c \frac{d t}{t} \ t^b e^{-t}$. The solution eq.~(\ref{metric})
describes a spacetime interpolating a ``de Sitter belt" spacetime near the origin and a Kerr spacetime far away from the origin.
The most important property of this spacetime is that there is no curvature singularity. For more details,
see ref.~\cite{Nicolini2010}. When the commutative limit $\th \rightarrow 0$ is taken, $m(r)=M$ and consequently the solution eq.~(\ref{metric})
recovers the ordinary Kerr solution. Since the noncommutative parameter is much smaller than the square of the horizon of black holes,
we can always take this limit near the horizon and outside.
But for some special cases, such as mini black holes, the noncommutative effect cannot be neglected.
Therefore, it is still interesting to consider the noncommutative deformation.

The (outer) event horizon located at $r_+$ satisfying $\Delta (r_+) =0$ can be written explicitly as
\begin{equation}
r_+^2 - 2r_+ m(r_+)  + a^2 =0.\label{horizoneq}
\end{equation}
It is impossible to have an analytic expression for $r_\pm$ from eq.~(\ref{horizoneq}). By using the iteration method,
one can solve the equation and give the outer horizon up to the order of $\mathcal{O}\left(\frac{1}{\sqrt{\th}} e^{-\frac{r_0^2}{4\th}}\right)$,
\begin{equation}
r_+ = r_0 \left(1 - \frac{r_0}{2 (r_0 - M)} \frac{2 M}{\sqrt{\pi\th}} e^{-\frac{r_0^2}{4\th}}\right)
+ \mathcal{O}\left(\frac{1}{\sqrt{\th}} e^{-\frac{r_0^2}{4\th}}\right),\label{outerhorizon}
\end{equation}
where $r_0 = M+\sqrt{M^2-a^2}$ is the outer horizon of the ordinary Kerr black hole. The horizon area is
\begin{equation}
A = 4 \pi (r_+^2+a^2) = 4 \pi (r_0^2+a^2)-\frac{4 \pi r_0^3}{r_0-M} \frac{2 M}{\sqrt{\pi \th}} e^{-\frac{r_0^2}{4\th}}
+ \mathcal{O}\left(\frac{1}{\sqrt{\th}} e^{-\frac{r_0^2}{4\th}}\right).\label{area}
\end{equation}
The first term of the above equation is just the horizon area of the ordinary Kerr black hole. In the rest of our paper,
we shall discuss the thermodynamics of the noncommutative inspired Kerr black hole by means of the reformulated tunneling method.

\section{Radiation via quantum tunneling}
We consider a tunneling process of a massless scalar particle at the outer horizon,
which corresponds to a real scalar field in the background of the noncommutative inspired Kerr black hole.
Note that a dimensional reduction technique similar to that in ref.~\cite{Umetsu09}
can be utilized in the study of the noncommutative inspired Kerr black hole
although the geometry of the noncommutative inspired black hole is different from that of the commutative one.
With this technique, we reduce the action of the scalar field near the horizon to an effective (1+1)-dimensional action
from which the equation of motion for the scalar field $\phi$ can easily be derived~\cite{Iso2006,Umetsu09},
\begin{equation}
\left[\frac{1}{F(r)} \left(\p_t - i A_t\right)^2 - F(r) \p_r^2 - F'(r) \p_r\right] \phi=0, \label{KGequation}
\end{equation}
where
\begin{equation}
A_t = m \O(r),\qquad\O(r) = - \frac{a}{r^2 + a^2},
\end{equation}
and
\begin{equation}
F(r) = \frac{\D}{r^2 + a^2},\qquad\D = r^2 - 2 r m(r)  + a^2. \label{entrance}
\end{equation}
Here $m$ is the magnetic quantum number of the tunneling particle. Note that this is just the Klein-Gordon equation for a free scalar field
interacting with a $U(1)$ gauge field in the following (1+1)-dimensional spacetime metric
\begin{equation}
ds^2 = -F(r) dt^2+\frac{dr^2}{F(r)}.
\end{equation}

Now we apply the WKB ansatz to solve eq.~(\ref{KGequation}). Considering the following relation
\begin{equation}
\phi (r,t) = e^{-\frac{i}{\hbar} S(r,t)},\label{phiansatz}
\end{equation}
we rewrite eq.~(\ref{KGequation}) as
\begin{equation}
(\p_t S)^2 + i \hbar \p_t^2 S + 2 \hbar A_t \p_t S + \hbar^2 A_t^2 - F^2 (\p_r S)^2 - i \hbar F^2 \p_r^2 S - i \hbar F F' \p_r S=0.\label{KGequation1}
\end{equation}
Substituting the expansion of $S(r,t)$,
\begin{equation}
S(r,t) = S_0 (r, t) + \sum_i \hbar^i S_i (r,t),
\end{equation}
into eq.~(\ref{KGequation1}) and equating the coefficients of the same powers of $\hbar$ on the both sides,
we obtain the first two equations corresponding to powers $\hbar^0$ and $\hbar^1$, respectively,
\begin{eqnarray}
&&\p_t S_0 = \pm F \p_r S_0, \label{0stequation}\\
&&2 \p_t S_0 \p_t S_1 + i \p_t^2 S_0 + 2 A_t \p_t S_0 - 2 F^2 \p_r S_0 \p_r S_1 - i F^2 \p_r^2 S_0 - i F F' \p_r S_0 =0.\label{1stequation}
\end{eqnarray}
As done in ref.~\cite{Banerjee09}, the solution of the order $\hbar^0$ equation (eq.~(\ref{0stequation})) is
\begin{equation}
S_0 (r, t) = \om \,(t \pm r_\star),\qquad r_\star (r) = \int^r \frac{d \ti{r}}{F(\ti{r})},\label{1stsolution}
\end{equation}
where $\om$ is given by
\begin{equation}
\om = E - m \O(r_+).
\end{equation}
Here $E$ is the conserved quantity corresponding to a timelike Killing vector and $\O(r_+)$ is the angular velocity evaluated at the horizon.
This $\om$ is identified as the effective energy experienced by the particle at the asymptotic infinity.

With the help of eq.~(\ref{0stequation}), we reduce the order $\hbar^1$ equation (eq.~(\ref{1stequation})) to be
\begin{equation}
\p_t S_1 + A_t = \pm F \p_r S_1. \label{1stequation1}
\end{equation}
Considering eq.~(\ref{1stsolution}), i.e. the solution of eq.~(\ref{0stequation}), we choose an ansatz for $S_1(r,t)$ as
\begin{equation}
S_1 (r,t) = \frac{\b_1}{A} \left(\om t + \ti{S}_1 (r)\right),
\end{equation}
where $\beta_1$ is a constant to be determined. Here the coefficient $\frac{\b_1}{A}$ appears due to the requirement of dimensional matching,
which takes a simpler form $\frac{\b_1}{r_+^2}$ for spherically symmetric black holes, see refs.~\cite{Banerjee,Banerjeeetal} for more details.
Substituting the ansatz into eq.~(\ref{1stequation1}) and solving the equation, we have the solution for $S_1(r,t)$,
\begin{equation}
S_1 (r,t) = \frac{\b_1}{A}\, \om \,(t \pm r_{\star\star}),\qquad
r_{\star\star}=\int^r \frac{d \ti{r}}{F(\ti{r})} \left(1 + \frac{A}{\b_1 \om} A_t (\ti{r})\right).
\end{equation}
As a consequence, we obtain the solution for $S(r,t)$ up to the first order of $\hbar$,
\begin{equation}
S(r,t) = \om \,(t \pm r_\star) +\hbar\, \frac{\b_1}{A} \,\om \,(t \pm r_{\star\star}). \label{solution}
\end{equation}
In the limit $m\rightarrow 0$, we can verify that the solution eq.~(\ref{solution}) coincides with the results given
in refs.~\cite{Banerjee,Banerjeeetal} for non-rotating black holes.

For the sake of convenience in the discussions below, we introduce the following coordinates
\begin{eqnarray}
&&u \equiv t - r_\star + \hbar \frac{\b_1}{A} (t-r_{\star\star}),\nonumber\\
&&v \equiv t + r_\star + \hbar \frac{\b_1}{A} (t+r_{\star\star}),\label{v}
\end{eqnarray}
which are just the null tortoise coordinates in the semiclassical limit ($\hbar \rightarrow 0$).
Expressing eq.~(\ref{solution}) in terms of these coordinates which are
defined inside and outside the event horizon, respectively, and then substituting it into eq.~(\ref{phiansatz}),
we can obtain the left and right modes for the both sectors around $r_+$,
\begin{equation}
\begin{array}{lll}
(\phi^{(\rm L)})_{\rm in} = e^{-\frac{i}{\hbar}\om v_{\rm in}},& \qquad
(\phi^{(\rm R)})_{\rm in} = e^{- \frac{i}{\hbar} \om u_{\rm in}},& \qquad (r_+ - \epsilon<r<r_+),\\
(\phi^{(\rm L)})_{\rm out} = e^{-\frac{i}{\hbar}\om v_{\rm out}},& \qquad
(\phi^{(\rm R)})_{\rm out} = e^{- \frac{i}{\hbar} \om u_{\rm out}},& \qquad (r_+<r<r_+ + \epsilon),\label{modes}
\end{array}
\end{equation}
where $\epsilon$ is an arbitrary small parameter.

At present we consider the tunneling process of a virtual pair of particles created near inside the horizon.
One particle tunnels quantum mechanically through the horizon and will be observed at infinity
while the other goes towards the center of the black hole. To describe this process, it is necessary to
choose a coordinate system which can cover both sides of the horizon. The Kruskal-like coordinates are a good choice.
The Kruskal-like time ($T$) and space ($X$) coordinates inside and outside the horizon are defined~\cite{Umetsu09} as follows:
\begin{equation}
\begin{array}{ll}
T_{\rm in} = e^{\kappa (r_\star)_{\rm in}} \cosh (\kappa t_{\rm in}),&
\qquad X_{\rm in} = e^{\kappa (r_\star)_{\rm in}} \sinh (\kappa t_{\rm in}),\\
T_{\rm out} = e^{\kappa (r_\star)_{\rm out}} \sinh (\kappa t_{\rm out}),&
\qquad X_{\rm out} = e^{\kappa (r_\star)_{\rm out}} \cosh (\kappa t_{\rm out}),\label{Kruskal}
\end{array}
\end{equation}
where $\kappa \equiv \frac{F'(r_+)}{2}$ is the surface gravity. With the help of eqs.~(\ref{outerhorizon}) and (\ref{entrance}),
we work out the surface gravity approximately,
\begin{equation}
\kappa = \kappa_0 \left[1 + \frac{r_0^2 (r_0^2-M r_0- M^2)}{2 (r_0^2 +a^2) (r_0 - M)^2} \frac{2 M}{\sqrt{\pi\th}}
e^{-\frac{r_0^2}{4\th}}\right],\label{kappa}
\end{equation}
where $\kappa_0 = \frac{r_0-M}{r_0^2+a^2} = \frac{\sqrt{M^2-a^2}}{2 M (M+\sqrt{M^2-a^2})}$ is the surface gravity of the ordinary Kerr black hole.

When the inside and outside coordinates are connected by\footnote{The temporal contribution to the Parikh-Wilczek's WKB calculation of Hawking
temperature was first considered in ref.~\cite{single}.}
\begin{equation}
t_{\rm in} \rightarrow t_{\rm out} - i \frac{\pi}{2\kappa},\qquad
(r_\star)_{\rm in} \rightarrow (r_\star)_{\rm out} + i \frac{\pi}{2\kappa},\label{connect-relation}
\end{equation}
we find the corresponding relations between the inside and outside Kruskal-like coordinates:
$T_{\rm in} \rightarrow T_{\rm out}$ and $X_{\rm in} \rightarrow X_{\rm out}$.
As a result,
we obtain from eq.~(\ref{v}) and eq.~(\ref{connect-relation}) the relations\footnote{To derive the connection rules
eq.~(\ref{uv-connect}), we have to know the relation between $(r_{\star\star})_{\rm in}$ and $(r_{\star\star})_{\rm out}$.
As the connection rule for $r_\star$ is
\[
(r_\star)_{\rm in} \rightarrow (r_\star)_{\rm out} + i \frac{\pi}{2\kappa}
= (r_\star)_{\rm out} + i {\rm Im} \int_{r_{\rm in}}^{r_{\rm out}} \frac{d \ti{r}}{F(\ti{r})},
\]
it is reasonable to give the connection rule for $r_{\star\star}$
as follows:
\[
(r_{\star\star})_{\rm in} \rightarrow (r_{\star\star})_{\rm out} + i {\rm Im} \int_{r_{\rm in}}^{r_{\rm out}} \frac{d \ti{r}}{F(\ti{r})}
\left(1 + \frac{A}{\b_1 \om} A_t (\ti{r})\right) = (r_{\star\star})_{\rm out} + i \frac{\pi}{2\kappa} \left(1 + \frac{A}{\b_1 \om} A_t (r_+)\right).
\]
We can also have this result from another point of view: Because $A_t (r)$ is well-defined and finite near the horizon,
we deal with it as a constant valued at the horizon $r_+$ in the integrand, as a consequence, we obtain
\[
r_{\star\star}
=\int^r \frac{d \ti{r}}{F(\ti{r})} \left(1 + \frac{A}{\b_1 \om} A_t (\ti{r})\right) = \left(1 + \frac{A}{\b_1 \om} A_t (r_+)\right) r_\star .
\]
Therefore, the above connection rule can easily be retrieved.}
connecting the coordinates $u$ and $v$ that are
defined inside and outside the event horizon, respectively,
\begin{eqnarray}
&&u_{\rm in} \rightarrow u_{\rm out} - i \frac{\pi}{\kappa} - i \hbar \frac{\b_1}{A} \frac{\pi}{\kappa}
- i\hbar \frac{\pi}{2 \kappa \om} A_t(r_+),\nonumber\\
&&v_{\rm in} \rightarrow v_{\rm out} + i \hbar \frac{\pi}{2\kappa\om} A_t(r_+).\label{uv-connect}
\end{eqnarray}
Under these connection rules the inside and outside modes are finally found to be connected by the following relations:
\begin{eqnarray}
&&(\phi^{(\rm L)})_{\rm in} \rightarrow e^{\frac{\pi}{2 \kappa} A_t(r_+)} (\phi^{(\rm L)})_{\rm out},\nonumber\\
&&(\phi^{(\rm R)})_{\rm in} \rightarrow e^{-\left[\left(1+ \hbar \frac{\b_1}{A}\right) \frac{\pi \om}{\hbar \kappa}
+ \frac{\pi}{2 \kappa} A_t(r_+)\right]} (\phi^{(\rm R)})_{\rm out}.\label{phi-connection}
\end{eqnarray}

\section{Emission spectrum}
We now derive the emission spectrum through utilizing the density matrix technique~\cite{Banerjee09,Umetsu09} and the connection rules
eq.~(\ref{phi-connection}).
Consider $n$ pairs of virtual particles created near inside the horizon. Each pair is
described by the modes given in the first line of eq.~(\ref{modes}). Therefore, the physical state of the system, when observed from outside,
can be represented as
\begin{equation}
|\Psi\rangle = N \sum_n |n^{(\rm L)}_{\rm in}\rangle \otimes |n^{(\rm R)}_{\rm in}\rangle \rightarrow
N \sum_n e^{- \frac{n \pi \om}{\hbar \kappa'}} |n^{(\rm L)}_{\rm out}\rangle \otimes |n^{(\rm R)}_{\rm out}\rangle,
\end{equation}
where the corrected surface gravity is defined by $\kappa' \equiv \kappa \left(1+\hbar \frac{\beta_1}{A}\right)^{-1}$.
$N$ is the normalization factor which can be determined by the normalization condition $\langle\Psi|\Psi\rangle =1$,
\begin{equation}
N = \frac{1}{\left(\sum_n^{} e^{-\frac{2 n \pi \om}{\hbar \kappa'}}\right)^{\frac{1}{2}}}.
\end{equation}
One can calculate the above sum for both bosons ($n=0, 1, 2, 3, \cdots$) and fermions\footnote{For fermions, such as spin-$\frac{1}{2}$ fields,
each component of spinors satisfies the Klein-Gordon equation. Therefore the left and right modes have the same relations as eq.~(\ref{modes}),
and we can then discuss fermions in the same way as bosons.} ($n=0, 1$). The results are
\begin{equation}
N_{(\rm boson)} = \left(1- e^{-\frac{2\pi\om}{\hbar \kappa'}}\right)^{\frac{1}{2}},\qquad
N_{(\rm fermion)} = \left(1+ e^{-\frac{2\pi\om}{\hbar \kappa'}}\right)^{-\frac{1}{2}}.
\end{equation}
In the following we focus only on bosons because the analysis for fermions is similar.
For bosons the density matrix operator of the system is given by
\begin{equation}
\hat{\rho}_{(\rm boson)} \equiv |\Psi\rangle_{(\rm boson)} \langle\Psi|_{(\rm boson)} = \left(1- e^{-\frac{2\pi\om}{\hbar \kappa'}}\right)
\sum_{n,m} e^{-\frac{n \pi\om}{\hbar\kappa'}} e^{-\frac{m\pi\om}{\hbar\kappa'}} |n^{(\rm L)}_{\rm out}\rangle
\otimes |n^{(\rm R)}_{\rm out}\rangle\langle m^{(\rm R)}_{\rm out}|\otimes\langle m^{(\rm L)}_{\rm out}|.
\end{equation}
After tracing out the ingoing (left) modes, we obtain the density matrix operator expressed only by the outgoing (right) modes,
\begin{equation}
\hat{\rho}_{(\rm boson)} = \left(1- e^{-\frac{2\pi\om}{\hbar \kappa'}}\right) \sum_n e^{-\frac{2 n \pi\om}{\hbar\kappa'}}
|n^{(\rm R)}_{\rm out}\rangle\langle n^{(\rm R)}_{\rm out}|.
\end{equation}
Therefore, the average number of particles detected at the asymptotic infinity takes the form
\begin{equation}
\langle n \rangle_{(\rm boson)} = {\rm trace} (\hat{n} \hat{\rho}_{(\rm boson)}) = \left(1- e^{-\frac{2\pi\om}{\hbar \kappa'}}\right)
\sum_n n e^{-\frac{2 n \pi\om}{\hbar\kappa'}} = \frac{1}{e^{\frac{2\pi\om}{\hbar\kappa'}} - 1}.
\end{equation}
This is exactly the Bose distribution that corresponds to a black body spectrum with a corrected Hawking temperature
\begin{equation}
T_{\rm h} = \frac{\hbar \kappa'}{2\pi} = \widetilde{T}_{\rm H} \left(1+\hbar \frac{\beta_1}{A}\right)^{-1},\label{Hawking temperature}
\end{equation}
where $\widetilde{T}_{\rm H} = \frac{\hbar \kappa}{2\pi} = \frac{\hbar F'(r_+)}{4\pi}$ is the modified semiclassical Hawking temperature involving
noncommutative effects. Note that the magnetic quantum number of tunneling particles $m$ has no effects on the radiation temperature
although it affects the connection rules
eq.~(\ref{phi-connection}) through $A_t(r_+)$.
To see the effect of noncommutativity, we calculate the temperature $T_{\rm h}$ approximately
by using eqs.~(\ref{kappa}) and (\ref{Hawking temperature}). The result is
\begin{equation}
T_{\rm h} = T_{\rm H} \left[1 + \frac{r_0^2 (r_0^2-M r_0- M^2)}{2 (r_0^2 +a^2) (r_0 - M)^2} \frac{2 M}{\sqrt{\pi\th}} e^{-\frac{r_0^2}{4\th}}\right]
\left(1+\hbar \frac{\beta_1}{A}\right)^{-1},\label{Hawking temperature1}
\end{equation}
where the first factor $T_{\rm H} = \frac{\hbar}{2\pi} \frac{\sqrt{M^2-a^2}}{2 M (M+\sqrt{M^2-a^2})}$ is the standard Hawking temperature
for the ordinary Kerr black hole, the second factor involves the noncommutative effect and the third factor emerges from the first order
quantum correction beyond the semiclassical limit. Only in the commutative limit ($\th \rightarrow 0$) and semiclassical limit ($\hbar\rightarrow 0$)
can the standard Hawking temperature be reproduced. The similar analysis can be applied to fermions and
the same temperature as that of bosons can be obtained from Fermi distribution.

\section{Entropy spectroscopy}
By following the procedure utilized in ref.~\cite{Umetsu09}, we turn to derive the entropy spectrum for the noncommutative inspired Kerr black hole.
First, we calculate the average energy and the average squared energy of the radiation particles by using eq.~(\ref{phi-connection})
and the normalization condition $|\phi^{(\rm R)}_{\rm out}|^2=1$,
\begin{eqnarray}
\langle \om \rangle &=& \frac{\int_0^\infty \left(\phi^{(\rm R)}\right)^\ast_{\rm in} \om \left(\phi^{(\rm R)}\right)_{\rm in} d\om}
{\int_0^\infty \left(\phi^{(\rm R)}\right)^\ast_{\rm in} \left(\phi^{(\rm R)}\right)_{\rm in} d\om} = \b^{-1},\nonumber\\
\langle \om^2\rangle &=& \frac{\int_0^\infty \left(\phi^{(\rm R)}\right)^\ast_{\rm in} \om^2 \left(\phi^{(\rm R)}\right)_{\rm in} d\om}
{\int_0^\infty \left(\phi^{(\rm R)}\right)^\ast_{\rm in} \left(\phi^{(\rm R)}\right)_{\rm in} d\om}=2 \b^{-2},\label{om-average}
\end{eqnarray}
where $\b$ is the inverse of the radiation temperature, $\b \equiv \frac{1}{T_{\rm h}}$.
Next, we give the uncertainty detected in the energy $\om$,
\begin{equation}
\langle \D \om\rangle = \sqrt{\langle\om^2\rangle - \langle\om\rangle^2} = \b^{-1} = T_{\rm h}.
\end{equation}
When we identify~\cite{Umetsu09} this uncertainty as the lack of information in the black hole energy due to particle emission, and then
consider the first law of black hole thermodynamics, we have
\begin{equation}
T_{\rm h} (\D S_{\rm bh}) = \D \om.
\end{equation}
As a result, we finally obtain the loss of the black hole entropy due to particle emission,
\begin{equation}
\D S_{\rm bh} =1,
\end{equation}
which implies that the entropy of the black hole is quantized in units of the identity. Consequently,
the entropy spectrum is equispaced and given by
\begin{equation}
S_{\rm bh}(n) = n, \label{entropy}
\end{equation}
which is independent of the noncommutativity and quantum correction.
Consequently, our result extends the universality of the entropy quantum, that is, it is not only independent of black hole parameters
and higher-order quantum corrections~\cite{Umetsu09,Banerjeeetal},
but also independent of the noncommutativity of spacetimes.

\section{Conclusion}
In this paper, we discuss the thermal dynamics of
the noncommutative inspired Kerr black hole in terms of the reformulated tunneling method~\cite{Banerjee09}.
We focus our attention on the tunneling process of massless particles, which is described by a real
scalar field in the background of the noncommutative inspired Kerr black hole. Through using a dimensional reduction technique as done in
refs.~\cite{Umetsu09,Iso2006}, we investigate the quantum tunneling related to the noncommutative inspired Kerr black hole
by equivalently solving a (1+1)-dimensional Klein-Gordon wave equation
in $(t, r)$ spacetime.
As usual, we have to make a WKB ansatz in order to solve the wave equation.
To see the effect caused by the angular momentum of the radiation particle, we calculate the solution to the first order of $\hbar$ in the WKB ansatz.
In addition, by using the density matrix technique we also derive the emission spectrum. The results show that
the radiation temperature (eq.~(\ref{Hawking temperature1})) is not affected by the magnetic quantum number
of the tunneling particle $m$ although the connection
relations of the left or right moving modes between inside and outside horizon sectors are modified by $m$. However,
we can see from eq.~(\ref{Hawking temperature1}) that the radiation temperature is indeed modified by the noncommutativity of spacetime $\th$.
Only in the commutative limit ($\th \rightarrow 0$) and semiclassical limit ($\hbar\rightarrow 0$) can the standard Hawking temperature be reproduced.

Moreover, following the procedure proposed in ref.~\cite{Umetsu09}, we give the entropy spectrum
and show that the quantization of entropy is affected neither by $m$ nor by $\th$.
When the effects of both noncommutativity and quantum corrections are taken into consideration,
the well-known Bekenstein-Hawking formula $S_{\rm BH}=\frac{A}{4 \hbar}$ must be corrected.
Because of the complicated expression of the temperature, see eqs.~(\ref{Hawking temperature1}) and (\ref{area}),
it is hard to provide a concise relation between the entropy and the horizon area.
If we neglect the noncommutativity, i.e. taking the commutative limit $\th\rightarrow 0$, and consider only the effect of quantum corrections,
we derive the modified Bekenstein-Hawking formula up to the first order of $\hbar$ and obtain
the well-known logarithmic corrections to black hole entropy,
\begin{equation}
S_{\rm bh} = \frac{A}{4\hbar} + \frac{\beta_1}{4} \ln A.
\end{equation}
Due to the appearance of the logarithmic corrections, the area quantum is not equispaced~\cite{Umetsu09} based on the entropy quantum
(eq.~(\ref{entropy})).
Therefore, we also find in the case of the noncommutative inspired Kerr black hole that
the notion of the quantum of entropy is more natural than that of the quantum of area.

At last we mention that the parameter $\beta_1$ remains to be determined in the present paper. In ref.~\cite{Banerjeeetal},
this parameter can be fixed in terms of its relation with
the trace anomaly of the energy-momentum tensor for spherically symmetric black holes.
However, it is unclear whether that treatment is valid for spherically nonsymmetric black holes. We leave it for further considerations.

\section*{Acknowledgments}
Y-GM would like to thank the Kavli Institute for Theoretical Physics China for kind hospitality where part of the work is performed.
This work is supported by the National Natural
Science Foundation of China under grants No.11175090 and No.10675061, and by the Fundamental Research Funds for the Central Universities
under grant No.65030021.


\end{document}